\documentclass[tightenlines,aps,twocolumn,
showpacs,nofootinbib,floatfix]{revtex4}

\usepackage{float}
\usepackage{amsmath}

\newcommand{\be}{\begin{equation}}
\newcommand{\ee}{\end{equation}}
\newcommand{\ba}{\begin{eqnarray}}
\newcommand{\ea}{\end{eqnarray}}

\def\vec#1{{\mbox{\boldmath$#1$}}}
\def\ket#1{\vert #1 \rangle}
\def\bra#1{\langle #1 \vert}

\newcommand{\ep}{\epsilon}

\begin{document}
\begin{titlepage}

\begin{flushright}
\vbox{
\begin{tabular}{l}
 FTPI-MINN-15/49\\
 TTP16-001
\end{tabular}
}
\end{flushright}

\vspace{0.6cm}

\title{  Higgs boson decay to two photons and  the dispersion relations
}

\author{Kirill Melnikov}
\email{kirill.melnikov@kit.edu}
\affiliation{ Institute for Theoretical Particle Physics, Karlsruhe Institute of Technology, 
Karlsruhe, Germany}
\author{Arkady Vainshtein}
\email{vainshte@umn.edu}
\affiliation{William I. Fine Institute for Theoretical Physics and School of Physics and Astronomy,
University of Minnesota, Minneapolis, MN 55455, USA}

\begin{abstract}

\vspace{2mm}

We discuss the computation of the Higgs boson decay amplitude to two photons 
through the $W$-loop using  dispersion relations. 
The imaginary part of the form factor $F_W(s)$ that parametrizes  this decay   
is unambiguous in four dimensions. 
When it is used to calculate  the unsubtracted dispersion integral,  the 
finite result for the form factor 
$F_W(s)$ is obtained. However, the $F_W(s)$ obtained in this way  
differs by a  constant term  from the result  of a diagrammatic computation, 
based on  dimensional regularization.  It is easy to accommodate the missing constant by writing a once-subtracted 
dispersion relation for $F_W(s)$ but it is unclear {\it why} the subtraction needs to be done. 
The goal of this paper is to investigate this question in detail. We show that the 
correct constant can  be recovered within a dispersive approach in a number of ways that, however, 
either require an introduction of an ultraviolet regulator or unphysical degrees of freedom; 
unregulated and unsubtracted computations in the 
unitary gauge are insufficient, in spite of the fact that such computations  give a finite 
result.

\end{abstract}

\maketitle

\thispagestyle{empty}
\end{titlepage}

\section{Introduction} 

The decay rate of the Higgs boson to two photons through the $W$-loop was  computed in the literature at least thirteen  times 
\cite{egn,ikh,svvz,GWW1,GWW2,pit,CCNS,HTW,MZW,SZC,SVVZ12,Jegerlehner:2011jm,Christova:2014mea}. 
The recent flurry of activity around this process, important for understanding Higgs boson properties, 
 was  caused by the fact that the original computations 
of the $H \to \gamma \gamma$ decay rate \cite{egn,ikh,svvz}, 
performed almost forty years ago, were challenged in  Refs.~\cite{GWW1,GWW2}. 
Among the follow up computations~\cite{pit,CCNS,HTW,MZW,SZC,SVVZ12,Jegerlehner:2011jm,Christova:2014mea}, 
only  Ref.~\cite{Christova:2014mea} agreed with the findings of Refs.~\cite{GWW1,GWW2}. 

A good  way to describe the controversial situation is as follows.  
Consider the  $H \to \gamma(k_1) \gamma(k_2) $ decay amplitude, 
focusing on the $W$-boson loop, and write it as 
\be
{\cal M} = \frac{\alpha}{4 \pi v} \,F_W(m_H^2) (k_1^{\mu} \epsilon_1^{\nu}-k_1^{\nu} \epsilon_1^{\mu})(k_{2\mu}  \epsilon_{2\nu}-k_{2\nu}  \epsilon_{2\mu})\,.
\ee
Here $v\!=\!2m_{W}/g\!=\!\big(G_{F}\sqrt{2}\,\big)^{-1/2}$ is the Higgs  field vacuum expectation value
and $\ep_{1,2}$ are the photon polarization vectors.
The form factor $F_W(s)$ reads 
\be
\begin{split} 
& F_W(s)  = F_W^{\infty} + F_W^c (s),\\[1mm]
& F_W^c(s) =  3 \beta + 3\beta(2-\beta) f(\beta), 
\end{split} 
\label{eq3}
\ee
where $\beta = 4 m_W^2/s$ and 
\be
f(\beta) = -\frac{1}{4} \left [ \ln \frac{1+ \sqrt{1-\beta} }{1 - \sqrt{1-\beta} } - i \pi \right ]^2.
\label{fbeta}
\ee

The constant term $F_W^{\infty}$ 
in Eq.\,(\ref{eq3}) is the gist of the current discussion: 
according to  Refs.\,\cite{egn,ikh,svvz,pit,CCNS,HTW,MZW,SZC,SVVZ12,Jegerlehner:2011jm} $F_W^{\infty} = 2$ 
and according to Refs.\,\cite{GWW1,GWW2,Christova:2014mea},
$F_W^{\infty} = 0$.   The two groups  \cite{GWW1,GWW2,Christova:2014mea} that claim $F_W^{\infty}  = 0$ 
have used two different techniques in their computations that, however,  
have two   important features in common. Indeed, both groups  refuse to use 
the dimensional regularization,  so that all the algebraic manipulations are  performed 
in four dimensions  {\it and} both groups  insist on  using  
only physical degrees of freedom in their  
calculations, i.e. the unitarity gauge for the $W$-bosons. 

The authors of Refs.~\cite{GWW1,GWW2} do  this in the context of 
Feynman diagrams and loop integrations. This is a delicate matter since 
all the individual diagrams are divergent and need to be combined before 
the actual integration over the loop momentum 
to ensure the finite result.  It is understandable, that this method of calculation drew criticism 
from Refs.~\cite{MZW,SVVZ12,Jegerlehner:2011jm,WS}.
Interestingly, the authors of Refs.~\cite{GWW1,GWW2} recognize this issue  and 
try to ameliorate it by imposing an {\it additional}  requirement on their result. This requirement 
is the heavy Higgs boson decoupling condition  $F_{W}(s\to \infty)=F_W^{\infty} = 0$ whose validity was, 
however,   criticized  in Refs.~\cite{SVVZ12,MZW,Jegerlehner:2011jm,SZC}.
Indeed, the decoupling limit, $\beta=4m_{W}^{2}/m_{H}^{2} \to 0$, can also be viewed as the limit $m_{W}\to 0$. 
It is well-known  that in the $m_W \to 0$ limit  the Higgs boson 
interaction with vector bosons, $2Hm_{W}^{2}W^{\dagger}_{\mu}W^{\mu}/v$, does not vanish for
the longitudinal polarizations of the $W$ bosons.  This is in contrast 
to the Higgs interactions with  fermions that  do vanish in the zero fermion mass limit.

On the other hand, the computation of Ref.~\cite{Christova:2014mea}, based on the dispersive  approach, 
is well-grounded at first sight. If one wants to use the  four-dimensional 
set up and physical degrees of freedom, the best thing to do is to use 
dispersion relations for the form factor $F_W(s)$ whose imaginary part can be 
computed from tree-level Feynman diagrams. 
As it is seen from Eqs.\,(\ref{eq3}, \ref{fbeta}), ${\rm Im}\, F_{W}(s)$ 
does not depend on the ambiguity in $F_{W}^{\infty}$ and equals to
\be
{\rm Im} \,F_{W}(s)=\frac{3\pi}{2}\,\theta(1-\beta)\,\beta(2-\beta)\ln\frac{1+ \sqrt{1-\beta} }{1 - \sqrt{1-\beta}}\,.
\label{imfw}
\ee
Note, that the imaginary part does vanish in  the $\beta \to 0$ limit.

 The full function $F_W(s)$ is then reconstructed 
using the unsubtracted dispersion relation in $s$,
\be
F_W(s) = \frac{1}{\pi} \int \limits_{4m_W^2}^{\infty} 
\frac{{\rm d} s_1 {\rm Im} [ F_W(s_{1})] }{s_1 -s - i0}\,.
\label{unsub}
\ee
The result of the integration in Eq.\,(\ref{unsub}) 
is the form factor $F^{c}_{W}$ shown in Eq.\,(\ref{eq3}), which implies  that $F_{W}^{\infty} = 0$.  The authors 
of Ref.~\cite{Christova:2014mea} interpret this result as the supporting evidence for the computation reported in Refs.~\cite{GWW1,GWW2}.  However, it should be recognized that the use of the unsubtracted dispersion relation assumes  
that the form factor $F_{W}(s)$ vanishes at 
$s\to \infty$, i.e.  $F_{W}^{\infty} = 0$. In other words, decoupling is assumed, rather than proved 
in Ref.~\cite{Christova:2014mea}. 
Without such an assumption, one can just add any real constant
to the right hand side of Eq.\,(\ref{unsub}).  

The constant  $F_W^\infty$ then either needs to be computed with a method that is 
different  from the dispersion relations {\it or} 
one should have a physical argument that determines the value of the form factor $F_W(s)$ for {\it one} value of $s$.
The most well-known example of the latter is the requirement that the Dirac form factor of the electron equals 
to one  at zero momentum transfer.

In case of the form factor $F_W(s)$, the low-energy theorem of Ref.~\cite{svvz} fixes its value 
at $s = 0$ to be the $W$-boson contribution to the coefficient of the one-loop QED $\beta$-function  $b_W$
\be
\lim_{s \to 0} F_{W} =b_{W}=7\,.
\ee
It is straightforward to check, using Eq.\,(\ref{eq3}), that  this condition at $s=0$ 
implies that $F_{W}^{\infty}=2$.

{\it Nevertheless, we can ask under which conditions 
the  dispersion relations  {\it without} the integral over the infinitely remote contour and the subtraction 
constant can be used in  general}.  
The answer  to this question is well-known. Such a possibility should exist if a {\it finite} 
form factor is computed in a  {\it renormalizable} 
theory  since each independent subtraction term corresponds to an independent renormalization 
condition that usually are fixed by considering divergent, rather than finite, quantities. 
Also, the use of unsubtracted dispersion relations  should be possible 
if one combines an ultraviolet (UV) regularization, such as dimensional or Pauli-Villars,
with the dispersion relations. Indeed, taking the dimensional regularization as an example, any 
integral over the infinitely 
remote  integration contour  can be discarded since $F_W(s) \sim s^{-\ep}$ for dimensional reasons 
and $\ep$ can always chosen in such a way that such an integral vanishes.   
In case of the Pauli-Villars regularization,   
$F_W(s)$ is also decreasing for values of $\sqrt{s}$ that are larger than the ultraviolet  cut-off, 
given by the regulator mass $M_{\rm PV}$. 

Combining  these observations with 
the fact that the ${\rm Im}\,F_W(s)$ in Eq.\,(\ref{imfw}) is finite and integrable in the dispersion 
integral, and that the Standard Model is, obviously, a renormalizable theory, 
we  conclude  that  something unusual should occur in ${\rm Im}\,F_W(s)$
in the limit when the regulators  are taken to their limiting values ($\ep \to 0$ or $M_{\rm PV} \to \infty$).
 Indeed, as we will see, this is exactly what happens and an additional 
contribution  to the imaginary part of the form factor is generated 
at  $\sqrt{s}$ of the order of the ultraviolet cut-off. This additional contribution 
to the dispersion integral changes  $F_W(s)$  if  $s$ is in the range 
$m_W  \ll \sqrt{s} \ll M_{\rm PV} $,  effectively leading 
to a  non-vanishing ``constant''  contribution to  $F_W$.

Although this approach may look  somewhat unphysical because it refers to the 
behavior of the theory for values of  Higgs masses that are  larger than 
the UV cut-off of the theory, we will see that  it is  consistent 
 with an infrared condition, e.g.\ the fixed value of $F_{W}$ at $s = 0$.  
We investigate how this happens in detail in this paper. 

\section{Longitudinal Polarizations}
\label{sect:long}

The issue of non-decoupling at $m_{W}=0$ refers to the longitudinally-polarized   $W$ bosons.
To describe these polarizations at large energies, $E\gg m_{W}$, one can substitute $W_{\mu}=\partial_{\mu} \phi/m_{W}$
where $\phi$ is the charged scalar field; this statement is the essence of the 
 equivalence theorem \cite{eqv1,eqv2,eqv3}. 
When written in terms of $\phi$-fields, the interaction of the $W$-bosons with 
the Higgs field  $2(H/v)m_{W}^{2}W^{\dagger}_{\mu}W^{\mu}$ takes the form
\be
S_{\rm int} = \int {\rm d}^4 x \,\frac{H}{v}\,\partial_{\mu}\partial^{\mu} \big(\phi^{\dagger}\phi\big)\,.
\label{intH}
\ee
Technically, this interaction  looks as a dimension-five, i.e. non-renormalizable, 
operator. This fact alone should act like  a  warning sign for the application of unsubtracted 
dispersion relations,  even if the result of the computation turns  out to be finite. 

We will study the contribution of the 
$\phi$ particles to the form factor for the two-photon Higgs decay assuming that Higgs-$\phi$ 
interaction is given by Eq.\,(\ref{intH}) and  
 denoting their masses as $m_\phi$.  We will see that this toy model captures
all the  essential features  of the 
problem discussed in the Introduction.   The counter-part 
of the full form factor $F_W(s)$ of Eq.\,(\ref{eq3}) in our toy model  
is denoted by  $F_\phi(s)$. 

There are two ways to deal with the operator in Eq.\,(\ref{intH}). 
The first one is based on the observation that for the purpose of computing $H \to \gamma \gamma$ 
decay amplitude,  it is possible to improve the 
ultraviolet properties of the action in Eq.\,(\ref{intH}). To this end, 
we  integrate by parts in Eq.\,(\ref{intH}), use equations of motion for the Higgs particle
$\partial_{\mu}\partial^{\mu}H=-m_{H}^{2}H$ and obtain 
\be
S_{\rm int} = -\frac{m_H^2}{v}\int {\rm d}^4 x\, H\, \phi^{\dagger}\phi \,.
\label{intH1}
\ee
This transformation  makes the interaction between the Higgs and the $\phi$'s explicitly renormalizable 
and  guarantees that  an unsubtracted dispersion relations for suitably defined form factor 
should be applicable. 

To proceed further, we parametrize  the matrix element $\bra{\gamma\gamma}\phi^{\dagger}\phi\ket{0}$
as follows 
\be
\bra{\gamma\gamma}\phi^{\dagger}\phi\ket{0}=
-\Phi(s) \cdot\frac{\alpha}{4\pi}\,f_1^{\mu \nu} f_{2,\mu \nu}\,,
\label{phig}
\ee
where $f_i^{\mu \nu} = k_i^\mu \epsilon_i^\nu - k_i^\nu \epsilon_i^\mu$. 
The physical form factor is then $F_\phi(s)= m_H^2 \Phi(s) =  s\, \Phi(s)$.

The form factor 
$\Phi(s)$ at large $s=(k_{1}+k_{2})^{2}\gg m_{\phi}^{2}$  equals to  \cite{svvz,LS,KMY}
\be
\Phi(s) = \frac{2}{s}\,.
\label{phig1}
\ee
After multiplying Eq.\,(\ref{phig}) by  the ``coupling constant'' 
$m_{H}^{2}$, identifying $m_H^2$ with $s $ and taking  the $s \to \infty$ limit, 
we obtain  $ \lim_{s \to \infty} F_{\phi}(s) =2$,  
which reproduces the non-decoupling constant in Eq.\,(\ref{eq3}).

It is straightforward to  reproduce this result in the  dispersive approach. 
Indeed, by unitarity
the imaginary part of the $\bra{\gamma\gamma}\phi^{\dagger}\phi\ket{\,0}$
amplitude is
\be
\begin{split}
&2\,{\rm Im}\,\bra{\gamma\gamma}\phi^{\dagger}\phi\ket{0}\!=\!
\int {\rm d Lips}(p_1,p_2,K_{12})\,M_{\phi \phi }^{\gamma\gamma}
\,,
\end{split}
\label{imf}
\ee
where $ {\rm d Lips}$ denotes the element of standard Lorentz invariant phase space of two $\phi$ particles
with momenta $p_{1}$ and $p_{2}$ and $M_{\phi \phi}^{\gamma \gamma}$ is the amplitude of $\phi(p_{1})+\bar\phi(p_{2}) \to \gamma(k_{1})+\gamma(k_{2})$ annihilation,
\be
M_{\phi \phi }^{\gamma\gamma}=2e^{2}\Big\{(\ep_{1}\ep_{2})-\frac{(p_{1}\ep_{1})(p_{2}\ep_{2})}{(p_{1}k_{1})}-\frac{(p_{1}\ep_{2})(p_{2}\ep_{1})}{(p_{1}k_{2})}\Big\}.
\ee
After integration over the phase space of two $\phi$ particles, we obtain 
\be
{\rm Im}\,\Phi(s) = - \pi \theta(1-\beta_\phi) \,
\frac{\beta_\phi}{s}\ln \frac{1+ \sqrt{1-\beta_\phi} }{1 - \sqrt{1-\beta_\phi}}\,.
\ee
We use this result in the unsubtracted dispersion relation and find 
\be
\Phi(s) =  \frac{1}{\pi} \int \limits_{4m_\phi^2}^{\infty} 
\frac{{\rm d} s_1 {\rm Im} \,\Phi(s_{1}) }{s_1 -s - i0}=\frac{2}{s}\,\big(1- \beta_\phi  f(\beta_\phi) \big)\,,
\label{eq14}
\ee
where $\beta_{\phi}\!=\!4m_{\phi}^{2}/s$. This expression coincides with Eq.\,(\ref{phig}) at large $s$ but it is 
valid for all $s$. To obtain the form factor $F_\phi(s)$, we multiply the real and imaginary parts 
of $\Phi$ 
by  $m_H^2$ and  identify $m_H^2$ with $s$. We find
\be
\begin{split}
& F_\phi(s) = s \Phi(s),\\[1mm]
&{\rm Im}\,F_{\phi}(s)=-\pi \,\theta(1-\beta_\phi)\beta_\phi 
\ln \frac{1+ \sqrt{1-\beta_\phi} }{1 - \sqrt{1-\beta_\phi}}\,,\\[1mm]
&F_{\phi}(s) = 2\big(1-\beta_\phi f(\beta_\phi)\big).
\end{split}
\label{flf}
\ee
Note that  the physical form factor $F_\phi(s)$ contains the constant 
contribution in the limit $s \to \infty$ and, therefore, does not support the 
$s \to \infty$ decoupling condition.

We can now ask what is the dispersion relation that the form factor $F_\phi(s)$ satisfies, 
provided that $\Phi(s)$ satisfies an unsubtracted dispersion relation. 
It is straightforward to answer this question. We  start  from 
the unsubtracted relation for $\Phi(s)$ in Eq.\,(\ref{eq14}), write  $\Phi = F_{\phi}/s$, and  obtain 
\be
F_{\phi}(s)=\frac{s}{\pi}\int_{4m_{\phi}^{2}}^{\infty} 
\frac{ds_{1} {\rm Im}\,F_{\phi}(s_{1})}{s_{1}(s_{1}-s-i0)}\,,
\label{eq15}
\ee
which is a once-subtracted dispersion relation for the form factor $F_\phi(s)$\,. Therefore, 
the  subtraction  of the dispersion relation for $F_\phi(s)$ 
at  $s\!=\!0$, which enforces  the condition $F_\phi(s=0)\!=\!0$,
 appears {\it automatically}  provided that we use the unsubtracted dispersion relations only for quantities 
(e.g. $\Phi(s)$)
that are computed in a theory where all  interactions 
are renormalizable by naive power-counting. This is {\it not}  the case for both, 
the toy model with the interaction term as in  Eq.\,(\ref{intH})  
and the Standard Model in the unitary gauge, so that 
 the use  of the  unsubtracted dispersion relations in both of these cases leads to incorrect results. 

We elaborate on the last statement. 
Suppose  that we do  not perform the integration by parts in the interaction term Eq.\,(\ref{intH})
and  use it directly to compute $H \to \gamma \gamma$ amplitude. Roughly speaking, this is a 
situation that 
corresponds to calculations in the unitary gauge in the full Standard Model. 
The imaginary part of this amplitude is given  
by the imaginary part of the physical form factor $F_\phi(s)$. If we now use this  imaginary part 
in the unsubtracted dispersion relation, we obtain a result that differs from $F_\phi(s)$ 
in Eq.\,(\ref{eq15}) by a subtraction constant 
\be
- \int \frac{{\rm d}s_1}{s_1} {\rm Im} [ F_\phi(s_{1})] = 2.
\label{unsub1}
\ee



We will now check that we can get the correct result for the form factor using the unsubtracted 
dispersion relations even if  we work 
with the non-renormalizable interaction in Eq.\,(\ref{intH}) but regulate the theory in the ultraviolet,   
in spite of the fact that the final result turns out to be finite.   

A simple form of the UV regularization is an introduction of  
Pauli-Villars fields. In our case it means that a contribution of the loop of charged scalar 
particles with the mass   $m_{\rm PV}$ should be subtracted from the loop of $\phi$-fields. 
The introduction
of the Pauli-Villars regulator leads to a change in the imaginary 
part of the form factor   ${\rm Im}\,F_{\phi}(s)$ at $s\ge 4m_{\rm PV}^{2}$\,,
\be 
\Delta_{\rm PV}[ {\rm Im}\,F_{\phi}] 
=\pi \,\theta(1-\beta_{R})\beta_{R} \ln \frac{1+ \sqrt{1-\beta_{R}} }{1 - \sqrt{1-\beta_{R}}}\,,
\ee
where $\beta_{R}=4m_{\rm PV}^{2}/s$. We find
\be
\Delta_{\rm PV}F_{\phi}(s)\!=\!\frac{1}{\pi} \!\!\int \limits_{4m_R^2}^{\infty} \!
\frac{{\rm d} s_1 \Delta_{\rm PV}[{\rm Im} [ F_\phi(s_{1})] }{s_1 -s - i0}\!=\!2\beta_{R}f(\beta_{R})\,.
\label{PV}
\ee
We are interested in the limit $\beta_{R}\!=\!4m_{\rm PV}^{2}/s \to \infty$; in that limit  
\be 
\Delta_{\rm PV}F_{\phi}(s) = 2\beta_{R}f(\beta_{R})\to 2\,, 
\ee
which is the same  constant that appears in Eq.\,(\ref{unsub1}). 

We will now  demonstrate that the same result is obtained if  dimensional regularization 
is used for the UV cut-off.
It is convenient to choose the photon polarization vectors as 
\be
\begin{split} 
& \epsilon_{1,2} =  \left ( 0, 1, \pm i, 0 \right )/\sqrt{2}\,,
\;\;\;\;
\epsilon_1 \cdot \epsilon_2 = -1,
\end{split}
\ee
in the reference  frame where the photon momenta are along $z$ axis.
Then from the unitarity relation 
\be
2\,{\rm Im}\, \langle\gamma\gamma  | H\rangle\!=\!
\int {\rm d Lips}(p_1,p_2,K_{12})\langle\phi\bar\phi \,| H\rangle \,M_{\phi \phi}^{\gamma\gamma}
\ee
we obtain 
\be
{\rm Im}\,F_{\phi}\!=\!(4\pi)^{2}\mu^{2\ep}\!\!\int\! \! {\rm d Lips}(p_1,p_2,K_{12})\Big\{\!-1\!+\frac{p_{x}^{2}+p_{y}^{2}}{p_{0}^{2}-p_{z}^{2}}\Big\},
\ee
where $\mu$ is the  normalization point and  the factor $\mu^{2\ep}$ restores a correct dimension.

At $d=4$ this expression shows that ${\rm Im}\,F_{\phi}\propto m_{\phi}^{2}$ and leads to
${\rm Im}\,F_{\phi}$ 
given in Eq.\,(\ref{flf}).\ At $d=4-2\ep$ we should split the $\phi$ particle 
momentum $p_{\mu}$ into the  four-dimensional part and the part $\vec p_{\ep}$ living in remaining
$-2\ep$ dimension. To determine an additional part $\Delta_{\ep}{\rm Im}\,F_{\phi}$ we put $m_{\phi}=0$. Then, 
\be
\begin{split}
\Delta_{\ep}{\rm Im}&\,F_{\phi}\!=\!(4\pi)^{2}\mu^{2\ep}\!\!\int\!  {\rm d Lips}(p_1,p_2,K_{12})\,\frac{{\vec n}_{\ep}^{2}}{\sin^{2}\theta}\\
&=\ep\,(4\pi)^{2}\mu^{2\ep}\!\!\int\!  {\rm d Lips}(p_1,p_2,K_{12})\\
&=\frac{\Omega^{(d-1)}}{2^{d-3} (2\pi)^{d-4}}\,\ep\, \Big(\frac{s}{\mu^{2}}\Big)^{-\ep}
\approx 2\pi \ep\, \Big(\frac{s}{\mu^{2}}\Big)^{-\ep}\,.
\end{split}
\ee
In 
these equations ${\vec n}_{\ep}={\vec p}_{\ep}/|{\vec p}|$, the angle $\theta$ is between $\vec p$ and $\vec k$,
and $\Omega^{(d-1)}$ is the solid angle in $d-1$ spatial dimensions.
The correction to the imaginary part induces the following change in  $F_{\phi}$ 
\be
\Delta_{\ep}F_{\phi}(s)\!=\!\frac{1}{\pi} \!\!\int \limits_{4m_R^2}^{\infty} \!
\frac{{\rm d} s_1 \Delta_{\ep}{\rm Im} [ F_\phi(s_{1})] }{s_1 -s - i0}\!=\!2\,\Big(\!\!-\frac{s}{\mu^{2}}\Big)^{-\ep}.
\label{ep1}
\ee
If we consider value of $\sqrt{s}$ that are much smaller  
than  the scale $\mu\exp(1/(2\ep))$, which plays a role of the UV cut-off, $\Delta_{\ep} F_{\phi}(s)$ 
adds  the required  constant $2$ to $F_{\phi}(s)$, that is  reconstructed from the unsubtracted and unregulated 
dispersion relation.  

By considering the toy model for the interaction of the longitudinally-polarized 
gauge bosons with the Higgs boson, 
we showed that the reason for the 
appearance of the constant contribution to the $H \to \gamma \gamma$ form factor is  
the fact that  interactions between the Higgs boson and the electroweak bosons in the 
unitary gauge are not renormalizable by power counting. The correct result can be obtained by either introducing 
explicit ultraviolet regulator, in spite of the fact that the computation of the form factor 
leads to a finite result, or 
by switching to a formulation of the theory where interactions are  renormalizable
by power counting.    We also note  that 
the UV regularization leads automatically to a result that is consistent with 
the low-energy constraint  which   is $F_{\phi}(s=0)=0$ in our
 toy model . As we saw, imposing this condition was sufficient for the dispersive reconstruction. 
All of these approaches can be used to compute the complete form factor $F_W(s)$ in the dispersive 
approach;   in the next Section, we will do that by performing the dispersive computation 
in a renormalizable $R_\xi$ gauge and studying if the unitary gauge result is recoverd  in the  $ \xi \to \infty$ 
limit. 

\section{Imaginary part and the renormalizable gauge} 

Our goal is to compute the form factor $F_W(s)$ using unsubtracted and unregulated 
dispersion relations. As we have seen,  this requires a formulation of the theory 
where renormalizability is  apparent.  Hence, we are forced to consider the $R_\xi$ gauges.  

Similar to what has been done before, we will calculate the form factor using 
dispersion relations; for this we will need to compute its  imaginary part 
for  $s \ne m_H^2$.  It is important to recognize that the amplitude that 
describes the transition of  the off-shell Higgs  to two photons becomes 
gauge-dependent; this applies to the dependence of the imaginary part on 
the electroweak gauge parameter $\xi$ as well as to the loss 
of the transversality of the electromagnetic current.  

The second problem is easy to avoid by choosing the non-linear $R_\xi$ gauge where the electromagnetic 
gauge invariance is explicitly maintained. To this end, we can use 
\be
{\cal L}_{\rm gauge} = -\frac{1}{\xi} \left | D_\mu W^\mu - i \xi  m_W \phi \right |^2,
\ee
as the gauge fixing term with  $D_\mu = \partial_\mu - ie A_\mu$.  If we choose this gauge, some 
Feynman rules of a linear $R_\xi$ gauge get modified but this is not important for us. The important 
point is that the gauge-fixing term ${\cal L}_{\rm gauge}$ 
eliminates the $\phi W \gamma$ vertex. In addition, it is important for what follows that in  
the $R_\xi$ gauge  Lagrangian, linear or not,  the only interaction vertex that explicitly contains $m_H^2$ 
is the interaction vertex involving the Higgs boson and the two Goldstone $\phi$-fields
\be
{\cal L}_{H\phi^{\dagger}\phi} = -\frac{m_H^2}{v} H  \phi^{\dagger} \phi\,.
\label{hff}
\ee
The interaction of the $\phi$-fields with the photons are that of the scalar QED and follow from the 
Lagrangian  
\be
{\cal L}_{\phi} = |D_\mu \phi|^2.
\ee
The final remark that we need to make is that the mass squared of the Goldstone boson $\phi$ 
is $m_\phi^2 = \xi m_W^2$.

As we will now show this information is  all that we need to perform the computation of $F_W(s)$,  
given the results  that we already presented  in Section~\ref{sect:long}.
To facilitate the computation of the imaginary part of the form factor $F_W(s)$, we  use the already-mentioned 
fact that among many diagrams that contribute to the form factor,  the only 
interaction vertex that is proportional to $m_H^2$ comes from the $H \phi^+ \phi$ interaction term in Eq.\,(\ref{hff}).
Motivated by this  observation, we  write the imaginary part as the sum of two terms
\be
{\rm Im}[F_W^{R_\xi}(r_H,\beta,\xi)] = 
r_H G_1(\beta,\xi) + G_2(\beta,\xi)\,,
\ee
where $r_H = m_H^2/s$ and $\beta = 4m_W^2/s$.  The functions $G_{1,2}$ can be computed directly from Feynman 
diagrams, however this is not necessary. Indeed,  there is one constraint on the two functions that is available 
to us since if we compute the imaginary part for $s=m_H^2  $, we should recover the $\xi$-independent result 
for the imaginary part in the unitary gauge. This implies 
\be
G_1(\beta,\xi) + G_2(\beta,\xi)  = {\rm Im}[F^{c}_W(\beta)],
\ee
where ${\rm Im}[F^{c}_W(\beta)]$ is the imaginary part of the form factor in the unitary gauge defined 
in  Eq.\,(\ref{imfw}). Next, since  the only $m_H^2$-dependent term in the calculation of 
${\rm Im}[F_W^{R_\xi}(r_H,\beta,\xi)] $ comes from the diagrams with $\phi^{\dagger} \phi$\,-intermediate state, 
we can read off  $G_1$  from the 
imaginary part of the form factor $F_{\phi}$ in Eq.\,(\ref{flf}).
We obtain 
\be
G_1(\beta,\xi) =  {\rm Im}[F_\phi(\beta_{\phi}=\xi \beta)].
\label{eq1098}
\ee

We can use the above constraints to rewrite the 
 imaginary part of the form factor in a general $R_\xi$ gauge in a useful way. 
By adding and subtracting $G_1$, we find 
\be
  {\rm Im}[F_W^{R_\xi}(r_H,\beta,\xi)] \! =\! 
{\rm Im}[F_W^{c}(\beta)]\!+\!(r_H\!-\!1)  {\rm Im}[F_\phi(\xi \beta)]\!\,.
\label{eq1099}
\ee 
The second term here shows that it is the off-shell behavior that  differentiates  the singular
unitary gauge from the renormalizable $R_\xi$ gauge.

We can  now restore the real part of the form factor from its imaginary part using the 
{\it unsubtracted} dispersion 
relation for  $s = m_H^2$. The result of the calculation should be correct since the theory in 
$R_\xi$ gauge is renormalizable by power-counting.   To this end, we need to compute 
\be
F_W(m_{H}^{2}) = \frac{1}{\pi} \int \frac{{\rm d} s_1 \; {\rm Im}[F_W^{R_\xi}(r_H,\beta_1,\xi)]
}{ s_1 - m_H^2}\,.
\ee
To compute this integral, we use 
the expression for the imaginary part as in Eq.\,(\ref{eq1099}) and realize that 
the dispersion integral of ${\rm Im}[F_W^{c}(\beta)]$ reconstructs $F_W^{c}$, see Eq.\,(\ref{eq3}). 
We also  substitute $m_H^2 \to s$, to conform with the previous notations, 
and write the final result for the form factor as 
\be
\begin{split}
 F_W(s) & = 
F_W^{c}(s) - 
\frac{1}{\pi} \!\!\int \limits_{4  \xi m_W^2}^{\infty}\! \frac{{\rm d} s_1}{s_1} \,{\rm Im}[F_\phi(\xi\beta_1)]
\\
& = F_W^{c}(s) +2\,.
\label{eq2000}
\end{split}
\ee
We note that the integral over ${\rm Im}F_\phi$ in the above equation 
is $\xi$ independent and  coincides with  a similar integral in the toy model, see  Eq.\,(\ref{unsub1}).
In general, the above computation  shows that the form factors calculated 
in the $R_\xi$ gauge and the unitary gauge {\it differ by a constant}, 
related to the contribution of Goldstone bosons to the imaginary part of the $H \to \gamma \gamma $ 
amplitude. The mass of the Goldstone boson  $m_\phi^2 = \xi m_W^2$ remains arbitrary in the calculation, 
so that the limit $\xi \to \infty$ can be studied. It follows from Eq.(\ref{eq2000}) that 
 the Goldstone boson contribution to $F_W(s)$ 
{\it does not decouple} in the limit $\xi \to \infty$; this feature  leads to a difference between 
the results of the calculations in the unitary and the $R_\xi$ gauges. 
Finally,  the Goldstone boson  
contribution does not have a pole at $s = m_H^2$ and, therefore, does not contribute to the discontinuity 
of the form factor; for all practical purposes, it is a subtraction term.

\section{Conclusions}
In this paper, we discussed how the dispersion relation 
computation of the $H \to \gamma \gamma$ decay amplitude through 
the $W$-boson loop can be 
reconciled with the results of the diagrammatic computations that employ 
dimensional regularization.
As was pointed in  Ref.~\cite{Christova:2014mea},  if one computes the 
imaginary part of the 
form factor $F_W(s)$ in the four-dimensional space-time 
and then uses it in an  unsubtracted  dispersion integral to calculate  
the full form factor, 
one obtains the result that differs from the correct  one  by a constant term. 
The appearance of this  constant can be interpreted as the need to perform a subtraction in a finite 
dispersion  integral which is quite unusual.  

We have shown that the need to perform the subtraction in the dispersion  integral for form factors 
computed in the unitary gauge  is a consequence of the fact that the SM 
in the unitary gauge is not explicitly renormalizable. 
If  one regularizes the (apparently finite) 
calculation  by either introducing explicit UV regulator or starts from 
the formulation of the theory where the renormalizability is, in fact, apparent, 
one always obtains an additional contribution to the real part of the form factor. 
For values of $s$ below the ultraviolet cut-off, this contribution is, essentially, a constant 
and can be interpreted as the subtraction term in the dispersion relation. Unfortunately, unregulated  
and unsubtracted dispersion relation calculations, that employ 
unitary gauge,  do not seem to be sufficient even if they lead to finite results.

\vspace*{0.2cm}
\noindent
{\bf Acknowledgments}
K.M. would like to thank F.~Caola and L.~Tancredi for useful conversations.


\end{document}